# Karl Dragutin Rakos 1925 - 2011

*Andrew P. Odell* [A]

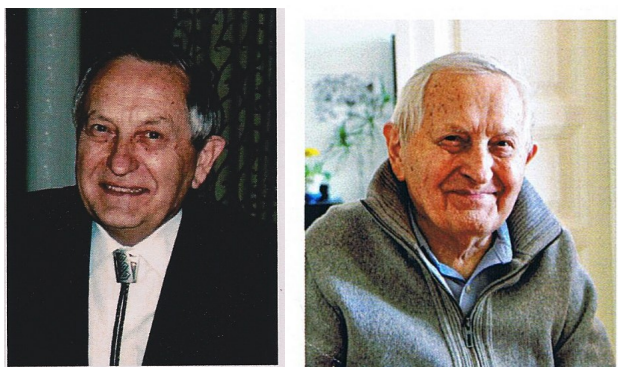

Professor Dr. Karl Dragutin Rakos passed away on October 31, 2011 one day before his 86th birthday.  With that the Vienna astronomical community lost a valued researcher, university teacher and co-founder of modern astrophysical research at the Institut für Astronomie of the  University of Vienna.

Karl "Drago" Rakos was born November 1, 1925 in Stefanje, Croatia into a family with two younger brothers during the very difficult economic times between the two World Wars. His interest in science and technology was evident even as a youth; he reminisced once that he first became aware of stars on Christmas eve when he was nine years old.  He began to read books in physics and chemistry, and much to his mother's dismay started experimenting with batteries in the kitchen.  By fourth grade he was building radios.

Later, Karl remembered the day he became captivated by the heavens – New Years Day of 1938.  He was bored and rifled through a box of books he found in the attic and ran across one on astronomy.  This naturally led to the study of science at the Universities of Zagreb and Belgrade.  Before the end of the war, the situation deteriorated drastically.  The approach of the Russian front brought communist terrorists, and the people prepared for the takeover.  He was forced to flee to avoid assassination, because he had expressed strong anti-communist views in school.  He was arrested and convicted in what was at that time Yugoslavia, so he could not finish his studies. Only after an adventurous flight in 1950 was he able to finish his studies in Graz.

In Graz he met his future wife, Annemarie (nee Gütschow) to whom he was happily married and they had four children, Christian, Imma, Anna, and Monika. In 1956, he graduated based on his work titled "Development and Construction of a Photoelectric Star Photometer".  His post doctoral work was as an assistant at the University Observatory in Graz and subsequently he held an assistant position at the University of Vienna Observatory.

---

[A]  Dept of Physics and Astronomy, Northern Arizona University Box 6010, Flagstaff AZ 86011 USA.
E-mail: Andy.Odell@nau.edu

Starting in 1960 with a Fulbright Scholarship (and later on an NSF grant), he spent four years in the United States doing pioneering research at Lowell Observatory in Flagstaff Arizona, where he met a young colleague, Dr. Alois Purgathofer, who also hailed from the-then University of Vienna Observatory. At that time, Harold Johnson was developing the UBV photometric system, and Karl used it to observe magnetic variable stars.

Rakos and Purgathofer immersed themselves in modern astrophysical research. At that time, the principal instrument at the University of Vienna Observatory was the Great Refractor, at one time the largest in the world, but by then more that 100 years old. Both astronomers must be credited with enthusiastically bringing astrophysical research to Vienna, and bringing Austria into the modern astronomical world..

Among other things, Rakos established highly advanced methods of digital electronics, for example, the development of a photon counting photometer and the introduction of digital detectors (at that time a Reticon, later CCDs). He was instrumental in the Institute acquiring the first computer processors, a PDP12 and PDP11, followed by a DEC-VAX, all of which led Rakos to introduce computerization to the Vienna Institute of Astronomy. He also developed the so-called Area Scanner, a devise used for precision astrometric and photometric measurements of double stars, which among other things led in 1974 to the first usable photometric data on the white dwarf companion of Sirius.

His work on chemically peculiar (Ap) stars, which he began at Lowell Observatory, was subsequently taken up by colleagues and continues to be a major research program at the Vienna Institute some 40 years later. That work also resulted in the first IAU Colloquium (Number 32) in Austria in 1975 on "Physics of Ap stars", with Karl D. Rakos as prominent supporter. It is no surprise, then, that in 1977 minor planet 4108 was named "Rakos".

Prof. Rakos headed the Institute of Astronomy from 1979 to 1981. In the years before and after, he directed his research objectives and goals toward modern astrophysical questions, which led to tensions with the those representing more traditional interests. In 1976, a report titled "A plan for astronomical research in Austria" was published, that he and a team of others developed for what was then the Federal Ministry of Science. The main elements of that plan are still in effect today, almost 40 years later, as the research goals of the astronomical institute at the University of Vienna.

I first met Karl when I was a visiting professor at the Institute in Vienna in the fall of 1985, very far from my home and friends. He invited us to dinner at his apartment for one Thursday afternoon; when we arrived, he and Annamarie had prepared a complete Thanksgiving dinner, with turkey and all the trimmings. That holiday, for Americans, is a time to get together with family and friends, and we realized that we had both, even though far from home. This is the kind of thoughtful person he turned out to be.

During the last several decades, his research interests shifted from questions about stars to the evolution of galaxies. The addition he made to this field was to use the Stromgren filter wavelengths to break the degeneracy of the UBV system to determine the age and metalicity of galaxies in clusters.  He also decided to use filters shifted in wavelength to match the redshift of the cluster, thus avoiding large k corrections.  In 1990 he began a twenty year collaboration with Jim Schombert, and in 1995 I joined the team.  His last research paper (Ages and Metallicities of Cluster Galaxies in A779 using Modified Strömgren Photometry) was published two months after his death by his final graduate student Yuvraj Harsha Sreedhar.

I spent many a night at the telescope with Karl, me being his final observing collaborator, and I remember those times very well.  He taught me how to obtain and reduce good photometry, and how to assess the quality of the data.  His minor, and then major stroke left him with less and less memory of the details of our work.  But the memories continue with me.

After retiring from the Institute in 1989, he became interested in his birth country again after 40 years.  He became a regular lecturer in Zagreb, and founded a chair of astronomy at the university.  He arranged for a 36-inch telescope owned by the Institute in Vienna to be transferred there and installed on the Island of Hvar.  The optics had come from the US Naval Observatory in Flagstaff, and the telescope had been built around that at the Vienna Institute.

Prof. Rakos was a member of the International Astronomical Union, specifically committee 26 (Double and Multiple Stars, serving as its president for six years), committee 27 (Variable Stars) and 42 (Close Double Stars). In addition, Prof. Rakos was a member of the American Astronomical Society, European Astronomical Society, Astronomische Gesellschaft, the European Science Foundation, the Croatian Astronomical Society, and an external member of the Croatian Acdemy of Sciences and Arts.  Prof. Rakos held collaborations with astronomers from around the world and observed at telescopes at Lowell Observatory, Kitt Peak National Observatory, United States Naval Observatory, and European Southern Observatory.

Above all this, his outstanding contributions as university teacher must be stressed since he supervised countless students and dissertations. Although the exact total of his "students" is hard to establish accurately, the more than two meter shelf space occupied in his office by black bound written works is most impressive. The high quality of Rakos' scientific achievements is demonstrated by the international network he enjoyed, encompasssing Europe, the USA, and ESO, AURA and ESA.

Prof. Karl Dragutin Rakos will be remembered by his friends and colleagues as an exemplary researcher and teacher, just as the minor planet bearing his name will continue to circle the sun for a long time.  A Gedächtnis-site has been set up for Karl: http://www.rudolf-albrecht.at/karl-rakos/

I would like to thank Karl's son Christian and Drs. Werner Weiss, Rudi Albrecht, Michael Breger, Otto Franz, and Tobias Kreidl for help to write this review, as well as Klaus Brasch for help in translating.